\documentclass[aps,prb,reprint,superscriptaddress,showpacs,longbibliography]{revtex4-2}

\usepackage{graphicx}
\usepackage{physics2}
\usephysicsmodule{ab}
\usepackage{amsmath,amsfonts}
\usepackage{tikz}
\usepackage{mathtools}
\usepackage[abs]{overpic}
\usepackage[justification=raggedright,singlelinecheck=false]{caption}
\usepackage{subcaption}
\usepackage{ulem}
\makeatletter
\newcommand\vb{\@ifstar\boldsymbol\mathbf}

\makeatother

\usepackage{hyperref}
\hypersetup{
	citecolor = blue,
	colorlinks = true,
	urlcolor = blue
}

\begin{document}

\title{
    Ferromagnetic resonance modulation in topological materials with bulk--boundary coexistence
    }

\author{Shun Muto}
\affiliation{Department of Physics, Nagoya University, Nagoya 464-8602, Japan}
\author{Yuya Ominato}
\affiliation{%
    Waseda Institute for Advanced Study, Waseda University, Shinjuku-ku, Tokyo 169-0051, Japan.
}%
\author{Takeo Kato}
\affiliation{Institute for Solid State Physics, The University of Tokyo, Kashiwa 277-8581, Japan}
\author{Mamoru Matsuo}
\affiliation{%
    Kavli Institute for Theoretical Sciences, University of Chinese Academy of Sciences, Beijing, 100190, China.
}%
\affiliation{%
    CAS Center for Excellence in Topological Quantum Computation, University of Chinese Academy of Sciences, Beijing 100190, China
}%
\affiliation{%
    Advanced Science Research Center, Japan Atomic Energy Agency, Tokai, 319-1195, Japan
}%
\affiliation{%
    RIKEN Center for Emergent Matter Science (CEMS), Wako, Saitama 351-0198, Japan
}%
\author{Ai Yamakage}
\affiliation{Department of Physics, Nagoya University, Nagoya 464-8602, Japan}

\date{\today}

\begin{abstract}
    We extend ferromagnetic resonance (FMR) modulation theory to describe systems in which bulk and boundary states of topological materials coexist, with both appearing at the same energy. As an application of the formulation, we investigate the enhancement of the Gilbert damping constant on the $\ab(110)$ surface of a $d$-wave superconductor where nodal quasiparticles coexist with edge states, which are one-dimensional boundary states, known as surface zero-energy Andreev bound states. We find two characteristic features: a pronounced edge-to-edge excitation peak near zero energy, and an additional edge-to-bulk excitation peak at the superconducting gap energy. We also observe power-law decay at low temperatures and exponential decay at intermediate temperatures in the low-energy regime. These features demonstrate the comparable contributions of the bulk and boundary states to the FMR response. Our theory provides a broadly applicable framework for the analysis of topological materials.
\end{abstract}

\maketitle

\section{Introduction\label{introduction}}
Measurement and control of magnetization dynamics are central to spintronics.
In particular, ferromagnetic resonance (FMR) in a ferromagnetic insulator (FI) interfaced with a conductor provides a sensitive probe of the spin dynamics in the adjacent fermionic system (FS) through interfacial exchange and spin pumping \cite{han2020spin}.
Under FMR modulation, the FS modifies both the resonance field and the linewidth in the microwave absorption spectrum of the FI.
Microscopically, these changes are governed by the real and imaginary parts of the dynamical spin susceptibility of the FS, and are often discussed in terms of the resonance-field shift $\Delta H$ and the enhancement of the Gilbert damping constant $\delta\alpha_{\mathrm G}$, which correspond to the peak shift and the linewidth broadening, respectively.
Such FI/FS-based FMR techniques have been applied to a broad class of fermionic systems including paramagnetic metals, atomic layer materials \cite{PhysRevB.87.140401, PhysRevB.94.205428}, superconductors \cite{PhysRevLett.100.047002, PhysRevB.104.144428}, and topological insulators \cite{PhysRevLett.113.196601, PhysRevLett.125.017204}.

Complementing these experiments, theoretical descriptions of FMR modulation have been developed over the past decade.
The phenomenological theory of spin pumping \cite{PhysRevLett.88.117601} and subsequent microscopic formulations \cite{PhysRevB.89.174417} established a relation between the FMR signal and the dynamical spin susceptibility of the FS, enabling material-specific predictions and extensions to a variety of systems \cite{doi:10.7566/JPSJ.89.053704, PhysRevB.96.024414, PhysRevB.99.144411, PhysRevB.106.144418, PhysRevB.104.054410, PhysRevB.107.174414, PhysRevB.105.205406, PhysRevB.106.L161406, PhysRevB.109.L121405, Ominato_2025}.
A key practical advantage of the microscopic approach is that it allows us to anticipate the characteristic frequency and temperature dependences of $\Delta H$ and $\delta\alpha_{\mathrm G}$, using an explicit model of the electronic structure.

Topological materials provide a test bed for the FMR modulation because their low-energy electronic states include topologically protected boundary modes and crucially affect interfacial spin dynamics.
In gapless topological phases, such as nodal topological superconductors and topological semimetals, gapless boundary states coexist with bulk states at the same energy \cite{PhysRevB.57.7997, PhysRevB.108.224503, Huang2015, doi:10.1126/science.aaa9297, Yano2023}. This bulk--boundary coexistence is expected to generate distinctive magnetic and transport responses, as the bulk and boundary states contribute comparably, leading to effects absent in either sector alone \cite{PhysRevB.93.245304, PhysRevB.93.235127, doi:10.1073/pnas.2313488121}.
Identifying clear dynamical signatures of such coexistence is therefore an important step toward characterizing topological materials by spin-based spectroscopies.

Despite the broad applicability of FMR modulation, the application of existing microscopic theories \cite{doi:10.7566/JPSJ.89.053704, PhysRevB.96.024414, PhysRevB.99.144411, PhysRevB.106.144418, PhysRevB.104.054410, PhysRevB.107.174414, PhysRevB.105.205406, PhysRevB.106.L161406, PhysRevB.109.L121405, Ominato_2025} to bulk--boundary coexisting systems is not straightforward.
Previous formulations \cite{doi:10.7566/JPSJ.89.053704, PhysRevB.96.024414, PhysRevB.99.144411, PhysRevB.106.144418, PhysRevB.104.054410, PhysRevB.107.174414, PhysRevB.105.205406, PhysRevB.106.L161406, PhysRevB.109.L121405, Ominato_2025} focus on either a translationally invariant bulk or a boundary state within an effective model. 
Consequently, these formulations describe only spin excitations within either the bulk or the boundary sector.
When bulk and boundary states coexist and both have finite spectral weight at the interface, however, an additional spin-excitation channel connecting the two sectors opens. This channel contributes to the FMR response, giving rise to a characteristic frequency dependence and distinct experimental signatures.
Therefore, a unified framework that directly probes the interfacial spin susceptibility of bulk--boundary coexistence in topological materials is required to analyze FMR modulation and enables disentangling the roles of bulk and boundary excitations in damping enhancement.

\begin{figure}
  \centering

  \begin{subfigure}{\columnwidth}
    \subcaption{}
    \includegraphics[width=0.8\columnwidth]{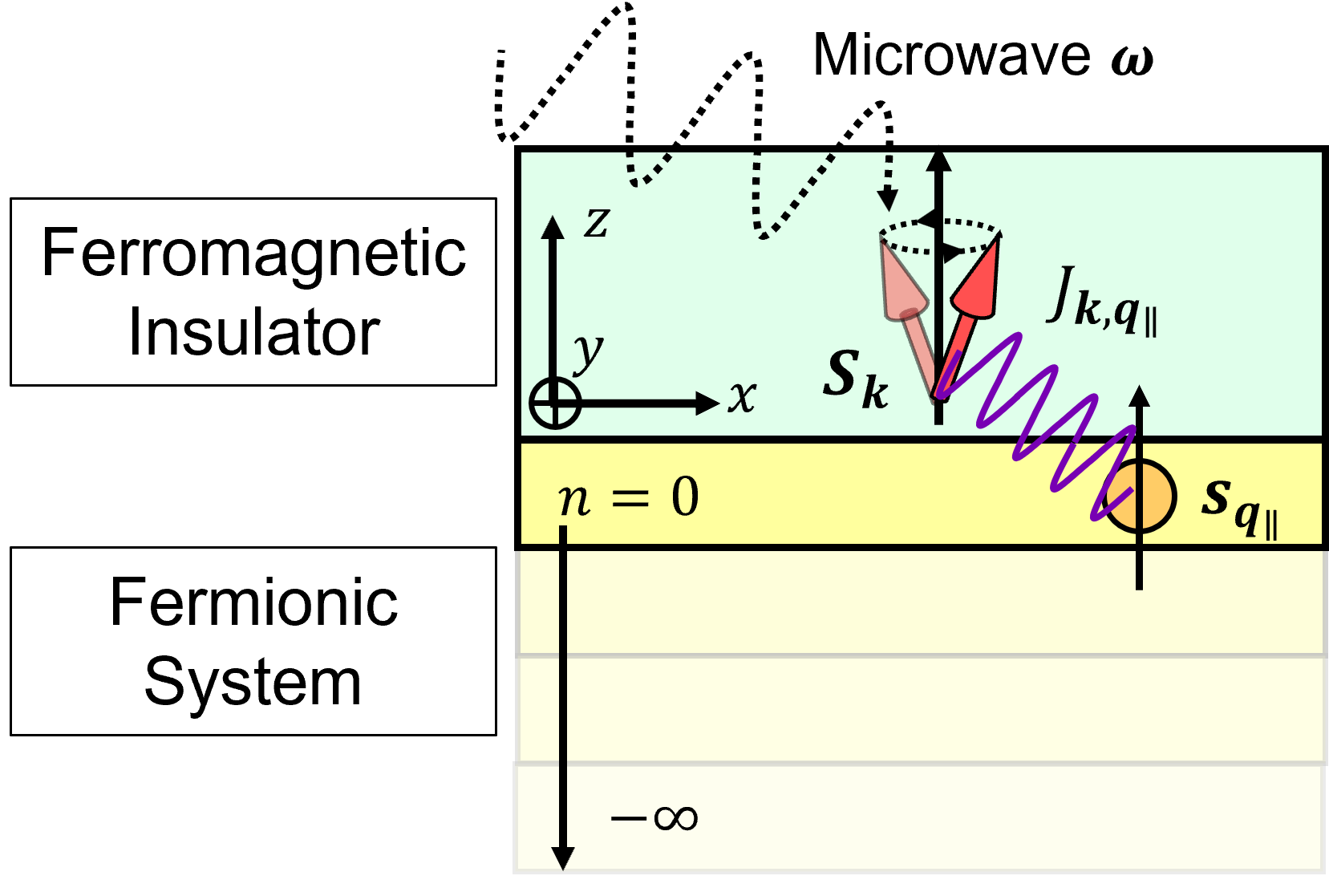}
    \label{system}
  \end{subfigure}

  \vspace{0.5em}

  \begin{subfigure}{\columnwidth}
    \subcaption{}
    \includegraphics[width=0.9\columnwidth]{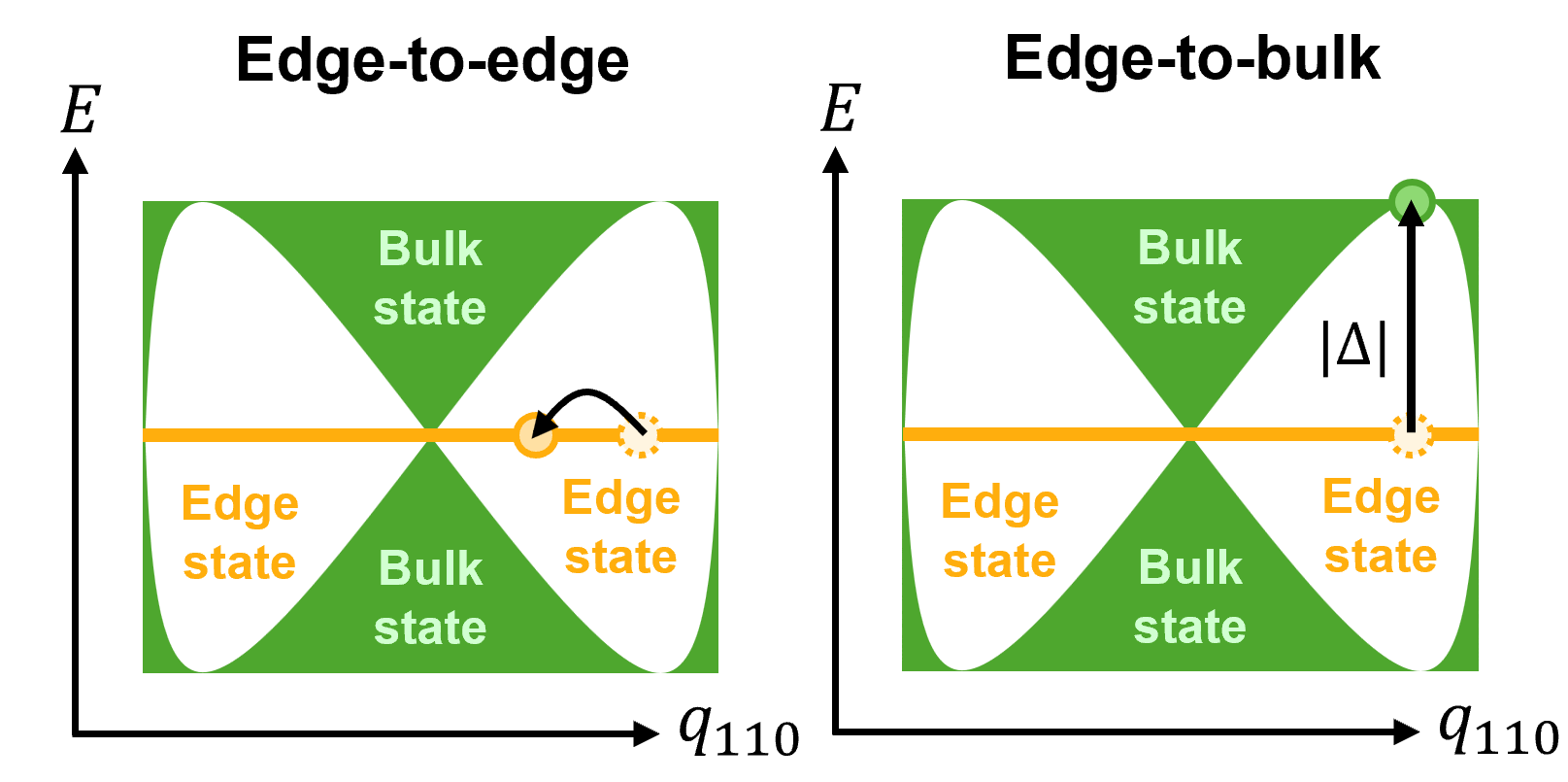}
    \label{excitation}
  \end{subfigure}
  
  \caption{(a) Schematic of the bilayer structure consisting of an FI and a semi-infinite FS. Microwave irradiation modulates the magnetization dynamics of an FI via interfacial spin interaction. (b) The illustration showing edge-to-edge excitation on the left and edge-to-bulk excitation on the right.}
\end{figure}

In this work, we develop a microscopic theory of FMR modulation for a bilayer composed of an FI and an arbitrary semi-infinite FS, as shown in Fig.~\ref{system}.
The magnon self energy, and hence $\Delta H$ and $\delta\alpha_{\mathrm G}$, are expressed in terms of the surface dynamical spin susceptibility of the FS.
Because surface susceptibility is computed from the surface Green’s function of a semi-infinite system, the formulation naturally includes both bulk and boundary states and considers their contributions on the same footing.
The resulting scheme is broadly applicable to multiband tight-binding models and provides an efficient route to quantitative calculations of FMR modulation in topological materials.
Moreover, the framework enables a microscopic analysis of how interfacial exchange coupling and interfacial disorder affect FMR modulation in systems with bulk--boundary coexistence. It also provides a basis for evaluating interfacial spin-transport parameters, such as the effective spin-mixing conductance, and, with further extensions, spin-transport length scales such as the spin-diffusion length.

As a concrete application, we study the (110) surface of a $d$-wave superconductor, one of the simplest settings where bulk nodal quasiparticles coexist with edge states, which are one-dimensional boundary states, known as surface zero-energy Andreev bound states (ABS)~\cite{PhysRevB.57.7997}.
We compute the frequency and temperature dependence of the enhanced Gilbert damping constant and identify two characteristic features on the (110) surface: (i) a pronounced peak (Fig.~\ref{excitation}, left) arising from quasiparticle excitation within the zero-energy surface bound states (edge-to-edge excitation), and (ii) an additional peak (Fig.~\ref{excitation}, right) near the superconducting gap scale associated with quasiparticle excitation from edge to bulk states (edge-to-bulk excitation).
These signatures demonstrate how bulk and boundary contributions can be comparable in FMR modulation and illustrate the usefulness of our framework for revealing bulk–-boundary coexistence.

This paper is organized as follows. In Sec.~\ref{model}, we introduce the model for the FI/FS bilayer and formulate the microscopic theory for a semi-infinite FS. In Sec.~\ref{d-wave}, we apply the theory to a $d$-wave superconductor surface and discuss the resulting frequency and temperature dependences of $\delta\alpha_{\mathrm G}$. Section~\ref{conclusion} summarizes our conclusions and perspectives.

\section{Microscopic Formulation\label{model}}
We consider the bilayer structure consisting of an FI and an arbitrary semi-infinite FS as shown in Fig.~\ref{system}. The Hamiltonian $H$ is written as
\begin{equation}
  H=H_{\mathrm{FI}}+H_{\mathrm{FS}}+H_{\mathrm{int}},
\end{equation}
where $H_{\mathrm{FI}}$, $H_{\mathrm{FS}}$, and $H_{\mathrm{int}}$ are the Hamiltonians of an FI, an arbitrarily semi-infinite FS, and the interaction between two materials at the interface, respectively. 
To isolate the essential aspect of the bulk--boundary coexistence effect, in this paper, we assume that the constituent materials, $H_{\rm FI}$ and $H_{\rm FS}$, are clean and take into account only the randomness of the exchange interaction $H_{\rm int}$ at the interface. 
Since the FMR modulation is an interfacial effect, incorporating such interfacial disorder is indispensable. Indeed, even if the materials themselves can be fabricated in an extremely clean manner, it is very difficult to realize a perfectly clean junction interface due to lattice mismatch and related factors; interfacial disorder should therefore be taken into consideration.

\subsection{Ferromagnetic Insulator (FI)}
We use a two-dimensional Heisenberg model for the FI because interfacial physics dominates the surface behavior. The Hamiltonian is written as 
\begin{equation}
  H_{\mathrm{FI}}=-\mathcal{J}\sum\limits_{\ab\langle  i,j\rangle}\vb{S}_{i}\cdot\vb{S}_{j}-\hbar\gamma  h_{\mathrm{dc}}\sum_{i}S_{i}^{z},
\end{equation}
where $\mathcal{J}$ is the coupling constant between the neighboring localized spins in FI, $\gamma$ is the gyromagnetic ratio, $h_{\mathrm{dc}}$ is the static magnetic field and $\vb{S}_{i}$ is the spin located at $\vb r_i$.
Employing the Holstein--Primakoff transformation \cite{PhysRev.58.1098} under spin-wave approximation:
\begin{subequations}
  \begin{align}
    S^{x}_{i}+iS^{y}_{i} & \equiv  S^{+}_{i}\simeq\sqrt{2S_{0}}b_{i},        \\
    S^{x}_{i}-iS^{y}_{i} & \equiv  S^{-}_{i}\simeq\sqrt{2S_{0}}b^{\dag}_{i}, \\
    S^{z}_{i}            & =S_{0}-b^{\dag}_{i}b_{i},
  \end{align}
\end{subequations}
and the Fourier transformation over the $N_{\mathrm{FI}}$ in-plane sites in the FI:
\begin{align}
  \begin{aligned}
    S_i^\pm = \frac{1}{\sqrt{N_{\mathrm{FI}}}}\sum\limits_{\vb{k}}
    e^{\pm i\vb{k}\cdot\vb{r}_{i}} S_{\vb k}^\pm,
  \end{aligned}\qquad\qquad \\
  b_{i}   =\frac{1}{\sqrt{N_{\mathrm{FI}}}}\sum\limits_{\vb{k}}e^{i\vb{k}\cdot\vb{r}_{i}}b_{\vb{k}},\quad
  b^{\dag}_{i}   =\frac{1}{\sqrt{N_{\mathrm{FI}}}}\sum\limits_{\vb{k}}e^{-i\vb{k}\cdot\vb{r}_{i}}b^{\dag}_{\vb{k}},
\end{align}
we get
\begin{align}
  H_{\mathrm{FI}}
   & = \sum\limits_{\vb{k}} \hbar\omega_{\vb k} b^{\dag}_{\vb{k}}b_{\vb{k}},
\end{align}
where $\hbar\omega_{\vb{k}}$ is the magnon dispersion.
The Green's function of the FI is given by
\begin{align}
  G^{R}_{\boldsymbol{k}}\ab(\omega)
   & = -\int_0^\beta d\tau e^{i \nu_m \tau}
  \ab\langle    S_{\vb{k}}^{+}\ab(\tau)S_{\vb{k}}^{-}(0)\rangle \biggr|_{i \nu_m \to \omega + i 0}
  \notag                                    \\&
  =\frac{2S_{0}/\hbar}{\omega-\omega_{\boldsymbol{k}}+i\alpha_{\mathrm{G}}\omega}.
\end{align}
Here, we introduce the Gilbert damping constant $\alpha_{\mathrm{G}}$ phenomenologically \cite{PhysRevLett.6.223, CHEREPANOV199381, 10.1063/1.5085922}.

\subsection{Semi-Infinite Fermionic System (FS)}
We consider a semi-infinite fermionic system that has translation symmetry along the in-plane directions and is described by a semi-infinite tight-binding model along the out-of-plane direction. The Hamiltonian is given by
\begin{equation}
  H_{\mathrm{FS}}
    =
  \sum_{\vb{q_{\parallel}}}
  \sum_{n=0}^{-\infty}
  \vb{c}^{\dag}_{\vb{q}_{\parallel},n}\vb{u}\vb{c}_{\vb{q}_{\parallel},n}
  +\sum_{\vb{q_{\parallel}}}
  \sum_{n=-1}^{-\infty}
  \ab(\vb{c}^{\dag}_{\vb{q}_{\parallel},n}\vb{t}\vb{c}_{\vb{q}_{\parallel},n+1}
  +\mathrm{H.c.})
  \label{FSHamiltonian}
\end{equation}
$\vb{c}_{\vb{q}_{\parallel},n}$ $(\vb{c}_{\vb{q}_{\parallel},n}^\dagger)$ is a fermion annihilation (creation) operator with in-plane wave number $\vb{q}_{\parallel}$ at the $n$th layer, $\vb{u}$ and $\vb{t}$ denote the onsite and hopping $N \times N$ Hamiltonians, respectively.
The retarded Green's function at the $n$th layer in the semi-infinite system is defined as
\begin{align}
  g^{R}_{\vb{q}_{\parallel},n}\ab(E)
  = -\int_{0}^{\beta} d\tau
  e^{i \epsilon_m \tau}
  \ab\langle    \vb{c}_{\vb{q}_{\parallel},n}\ab(\tau)\vb{c}_{\vb{q}_{\parallel},n}^{\dag}\ab(0)\rangle
  \biggr|_{i \epsilon_m \to E + i \delta}.
\end{align}
The retarded Green's function on the surface, $n=0$, can be obtained from the recursive Green's function method reported by A. Umerski \cite{umerski1997closed}, and the detailed derivation is written in Appendix~\ref{RGFcal}.

\subsection{Interfacial Exchange Interaction}
The third term $H_{\mathrm{int}}$ represents the exchange interaction at the interface. The $i$th FI spin $\vb{S}_{i}$ at $\vb{r}_{i}$ couples to the surface FS spin $\vb{s}_{j_\parallel,0}$ at $\ab(\vb{r}_{j_\parallel},n=0)$ via the in-plane exchange coupling $J_{i,j_{\parallel}}$.
The Hamiltonian is described as
\begin{align}\label{Hint}
  H_{\mathrm{int}}
   & =
  \sum_{i,j_\parallel} \frac{J_{i, j_{\parallel}}}{2}
  S_i^+ s_{j_{\parallel},0}^{-} +\mathrm{h.c.}
  \notag \\
   & =
  \sum\limits_{\vb{k},\vb{q}_{\parallel}}
  \frac{J_{\vb{k},\vb{q}_{\parallel}}}{2}S^{+}_{\vb{k}}s^{-}_{\vb{q}_{\parallel},0}+\mathrm{h.c.}
\end{align}
The Fourier expansion of the spin ladder operators is defined as
\begin{align}
  S_i^\pm               & = \frac{1}{\sqrt{N_{\mathrm{FI}}}}
  \sum_{\vb k}
  S_{\vb k}^\pm e^{\pm i \vb k \cdot \vb r_i},
  \\
  s_{j_\parallel,0}^\pm & =\frac{1}{N_{\parallel}}
  \sum_{\vb q_\parallel}
  s_{\vb q_\parallel,0}^\pm e^{\pm i \vb q_\parallel \cdot \vb r_{j_\parallel}},
  \\
  J_{\vb{k},\vb{q}_{\parallel}}
                        & =\frac{1}{N_{\parallel}\sqrt{N_{\mathrm{FI}}}}\sum\limits_{i,j_{\parallel}}J_{i,j_{\parallel}}
  e^{i\ab(\vb{k}\cdot\vb{r}_{i} - \vb{q}_{\parallel}\cdot\vb{r}_{j_{\parallel}})}.
\end{align}
$N_{\mathrm{FI}}$ and $N_{\mathrm{\parallel}}$ are the numbers of in-plane sites in the FI and FS, respectively. The spin operator $s^{\pm}_{\vb{q}_{\parallel},0}$ in the FS is given by
\begin{align}
  s^{\pm}_{\vb{q}_{\parallel},0}
  =\sum\limits_{\vb{p}_{\parallel}}\vb{c}^{\dag}_{\vb{p}_{\parallel},0}
  \frac{\sigma^{\pm}}{2}
  \vb{c}_{\vb{p}_{\parallel} \pm \vb{q}_{\parallel},0},
  \quad
  \ab({s}^{-}_{\vb{q}_{\parallel},0})^{\dag}={s}^{+}_{\vb{q}_{\parallel},0}, \\
  \begin{aligned}
    \vb{c}^{\dag}_{\vb{p}_{\parallel},0}=\ab(c^{\dag}_{\vb{p}_{\parallel}\uparrow,0},c^{\dag}_{\vb{p}_{\parallel}\downarrow,0}),\quad
    \vb{c}_{\vb{p}_{\parallel},0}=\,^{t}\ab(c_{\vb{p}_{\parallel}\uparrow,0},c_{\vb{p}_{\parallel}\downarrow,0}),
  \end{aligned}
\end{align}
where the $2 \times 2$ matrices are introduced as
\begin{align}
   & \begin{aligned}
       \sigma^{+}=
       \begin{pmatrix}
      0 & 2 \\ 0 & 0
    \end{pmatrix},\quad
       \sigma^{-}=
       \begin{pmatrix}
      0 & 0 \\ 2 & 0
    \end{pmatrix}
     \end{aligned}.
\end{align}

\subsection{Magnon self energy and FMR modulation}
Using a perturbation expansion with respect to $H_{\mathrm{int}}$, the Green's function of the FI can be extended to the bilayer system \cite{PhysRevB.105.205406, PhysRevB.106.L161406, Ominato_2025}
\begin{align}
  G^{R}_{\vb{k}}\ab(\omega)
   & =\frac{2S_{0}/\hbar}{\omega-\omega_{\vb k}+i\alpha_{\mathrm{G}}\omega-\ab(2S_{0}/\hbar)\Sigma^{R}_{\vb{k}}\ab(\omega)}\notag \\
   & =\frac{2S_{0}/\hbar}{\omega - \omega_{\vb k}-\gamma\Delta  H+i\ab(\alpha_{\mathrm{G}}+\delta\alpha_{\mathrm{G}})\omega},
\end{align}
where $\Delta  H$ and $\delta\alpha_{\mathrm{G}}$ are resonance-field shift and the enhanced Gilbert damping constant, respectively, and are described using the magnon self energy $\Sigma^{R}_{\vb{k}}\ab(\omega)$:
\begin{equation}
  \Delta  H=\frac{2S_{0}}{\gamma\hbar}\mathrm{Re}\ab[\Sigma^{R}_{\vb{k}}\ab(\omega)],\quad\delta\alpha_{\mathrm{G}}=-\frac{2S_{0}}{\hbar\omega}\mathrm{Im}\ab[\Sigma^{R}_{\vb{k}}\ab(\omega)].
\end{equation}
The second-order magnon self energy is given by
\begin{equation}
  \Sigma^{R}_{\vb{k}}\ab(\omega)=-\sum\limits_{\vb{q}_{\parallel}}\frac{|J_{{\vb{k}},\vb{q}_{\parallel}}|^{2}}{4}\chi^{R}_{\vb{q}_{\parallel}}\ab(\omega).
\end{equation}
$\chi^{R}_{\vb{q}_{\parallel}}\ab(\omega)$ is the surface dynamic spin susceptibility of the semi-infinite system defined as
\begin{equation}
  \chi^{R}_{\vb{q}_{\parallel}}\ab(\omega)=\int^{\beta}_{0}d\tau
  e^{i\nu_m\tau}\ab\langle   s^{+}_{\vb{q}_{\parallel},0}\ab(\tau)s^{-}_{\vb{q}_{\parallel},0}\ab(0)\rangle
  \biggr|_{i \nu_m \to \omega + i 0}.
  \label{eq:chi_q}
\end{equation}
Focusing on the uniform magnon mode: $\vb{k}=0$ and applying the configuration average to the in-plane coupling constant at the interface (see Appendix~\ref{conf.ave.} for details), we obtain
\begin{equation}
  \overline{\Sigma^{R}_{\vb{k}=0}\ab(\omega)}
  =-\frac{J^{2}_{1}}{N_{\mathrm{FI}}}\chi^{R}_{\vb{q}_{\parallel}=0}\ab(\omega)-\frac{J^{2}_{2}}{N_{\mathrm{FI}}}\frac{1}{N_{\mathrm{\parallel}}}\sum\limits_{\vb{q}_{\parallel}}\chi^{R}_{\vb{q}_{\parallel}}\ab(\omega),
\end{equation}
where $J_{1}$ and $J^{2}_{2}$ are the mean and variance strength of the interface coupling, respectively.

\section{Application to a d-wave superconductor}
\label{d-wave}
\begin{figure}
  \centering
  \includegraphics[width=0.9\linewidth]{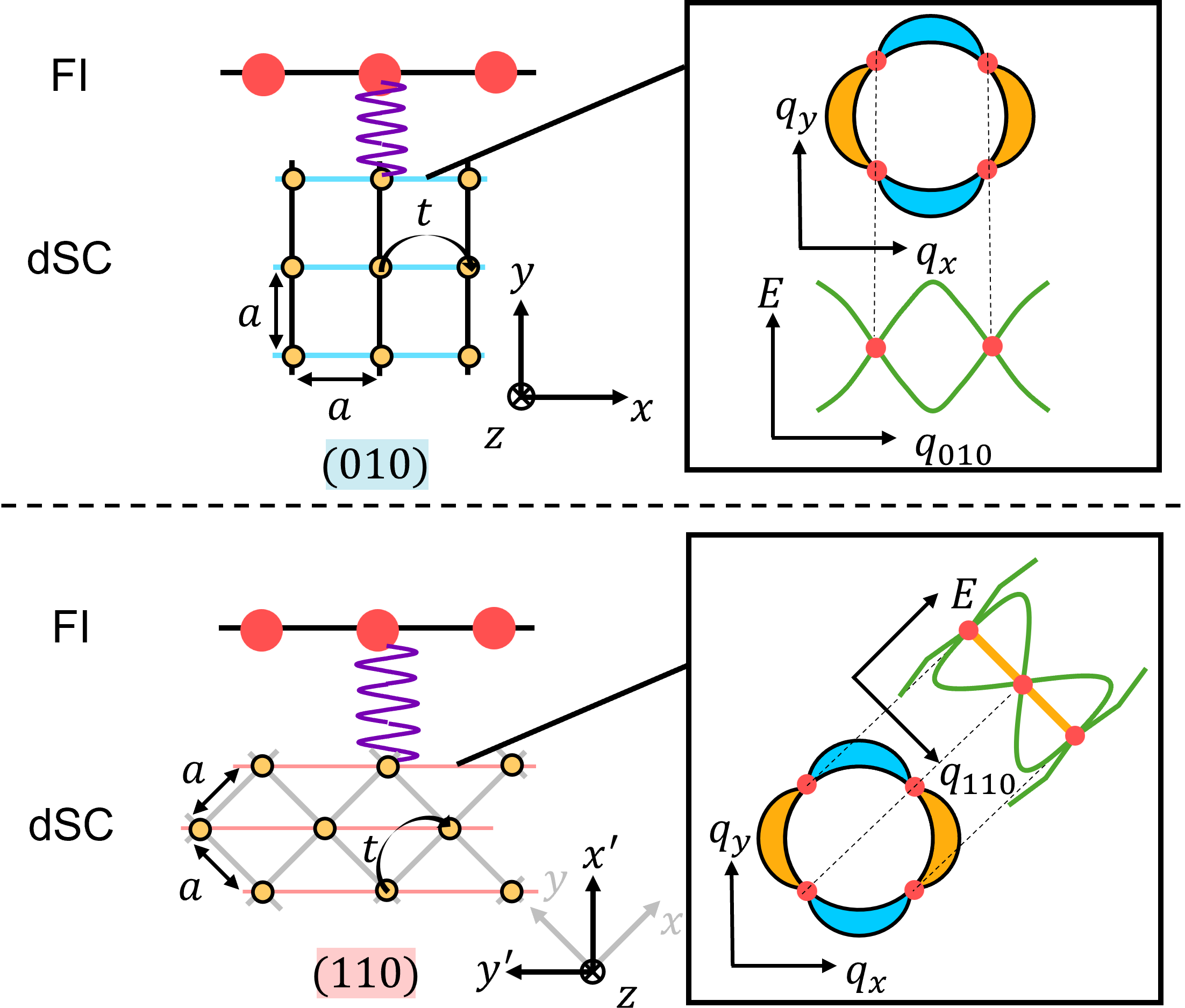}
  \caption{Left: schematic picture of the bilayer structure between an FI and a $\ab(010)/\ab(110)$ $d$-wave superconductor. Right: Fermi surfaces and energy spectra for the $\ab(010)/\ab(110)$ surface.}
  \label{dSCimageconnection}
\end{figure}

To clearly illustrate the essential features of our theory, we calculate $\delta\alpha_{\mathrm{G}}$ for two representative surfaces of a $d$-wave superconductor: the $(010)$ surface, which supports only bulk states, and the $\ab(110)$ surface, which also hosts edge states \cite{PhysRevB.57.7997}, as shown in Fig.~\ref{dSCimageconnection}.

\subsection{Surface Hamiltonian}
We use a two-dimensional square lattice model for the $d$-wave superconductor and assume the setup shown in Fig.~\ref{dSCimageconnection}. We define the Nambu basis as
\begin{equation}
  \vb{\tilde{c}^{\dag}}_{q,n}
  =\,\ab(\vb{c}^{\dag}_{q,n},\,^{t}\vb{c}_{-q,n}),\quad
  \vb{\tilde{c}}_{q,n}
  =\,^{t}\ab(\,^{t}\vb{c}_{q,n},\vb{c}^{\dag}_{-q,n}),
\end{equation}
and denote the lattice constant, nearest-neighbor hopping, chemical potential, and superconducting gap by $a, t, \mu$ and $\Delta$, respectively. The superconducting gap is given by
\begin{equation}
  \Delta=\Delta_{0}\tanh\ab(1.74\sqrt{\frac{T_{\mathrm{c}}}{T}-1}).
\end{equation}
For the $\ab(010)$ surface, the Hamiltonian is given by substituting $\vb u = \vb u_{010}(q_{010})$ and $\vb t = \vb t_{010}$ into Eq.~\eqref{FSHamiltonian}:
\begin{align}
    \label{010u}
    \vb{u}_{010}\ab(q_{010}) &=
    \begin{pmatrix}
      \xi_{q_{010}}\sigma^{0}     & \Delta_{q_{010}}\sigma^{y}
                                  \\
      -\Delta_{q_{010}}\sigma^{y} & -\xi_{q_{010}}\sigma^{0}
    \end{pmatrix},\\
    \label{010t}
    \vb{t}_{010} &=
    \begin{pmatrix}
      -t\sigma^{0}                & -i\frac{\Delta}{4}\sigma^{y}
                                  \\
      i\frac{\Delta}{4}\sigma^{y} & t\sigma^{0}
    \end{pmatrix},
\end{align}
where
\begin{equation}
    \xi_{q_{010}}=-2t\cos\ab(q_{010}a)-\mu,\quad
    \Delta_{q_{010}} = i\frac{\Delta}{2}\cos\ab(q_{010}a),
\end{equation}
and the Brillouin zone is $q_{010}\in\ab[-\pi/a,\pi/a]$. On the other hand, for the $\ab(110)$ surface, the Hamiltonian is obtained by substituting $\vb u = \vb u_{110}$ and $\vb t = \vb t_{110}(q_{110})$ into Eq.~\eqref{FSHamiltonian}:
\begin{align}
    \label{110u}
    \vb{u}_{110} &=
    \begin{pmatrix}
      -\mu\sigma^{0} & 0
                     \\
      0              & \mu\sigma^{0}
    \end{pmatrix},\\
    \label{110t}
    \vb{t}_{110}\ab(q_{110}) &=
    \begin{pmatrix}
      \xi_{q_{110}}\sigma^{0}     & \Delta_{q_{110}}\sigma^{y} 
                                  \\
      -\Delta_{q_{110}}\sigma^{y} & -\xi_{q_{110}}\sigma^{0}
    \end{pmatrix},
\end{align}
where
\begin{equation}
    \xi_{q_{110}}=-2t\cos\ab(\frac{q_{110}a}{\sqrt{2}}),\quad
    \Delta_{q_{110}}=\frac{\Delta}{2}\sin\ab(\frac{q_{110}a}{\sqrt{2}}),
\end{equation}
and the Brillouin zone is $q_{110}\in\ab[-\pi/\sqrt{2}a,\pi/\sqrt{2}a]$ \cite{PhysRevB.70.144502, PhysRevB.65.064507, PhysRevB.59.1421, PhysRevB.72.184510}.

\subsection{Gilbert damping  Enhancement}
  To compute $\delta\alpha_{\mathrm{G}}$, we use the fermionic Matsubara formalisms in the Gor'kov Green's function $\tilde{g}_{\vb{q}_{\parallel},n}\ab(i\epsilon_{m})$, which has the form:
  \begin{align}
    \tilde{g}_{\vb{q}_{\parallel},n}\ab(i\epsilon_{m})
     & =\int^{\beta}_{0}d\tau   e^{i\epsilon_{m}\tau}\ab(-\ab\langle\vb{\tilde{c}}_{\vb{q}_{\parallel},n}\ab(\tau)\vb{\tilde{c}}^{\dag}_{\vb{q}_{\parallel},n}\ab(0)\rangle)\notag \\
     & =\int^{\beta}_{0}d\tau   e^{i\epsilon_{m}\tau}
    \begin{pmatrix}
      g_{\vb{q}_{\parallel},n}\ab(\tau)       & f_{\vb{q}_{\parallel},n}\ab(\tau)
      \\[1ex]
      \bar{f}_{\vb{q}_{\parallel},n}\ab(\tau) & \bar{g}_{\vb{q}_{\parallel},n}\ab(\tau)
    \end{pmatrix}
    \notag                                                                                                                                                                         \\&
    =
    \begin{pmatrix}
      g_{\vb{q}_{\parallel},n}\ab(i\epsilon_m)       & f_{\vb{q}_{\parallel},n}\ab(i\epsilon_m)       \\[1ex]
      \bar{f}_{\vb{q}_{\parallel},n}\ab(i\epsilon_m) & \bar{g}_{\vb{q}_{\parallel},n}\ab(i\epsilon_m)
    \end{pmatrix}.
  \end{align}
  The normal $g$ and anomalous $f$ Green's functions in imaginary time are defined by \cite{Coleman_2015}
  \begin{align}
    g_{\vb{q}_{\parallel},n}\ab(\tau)
     & =-\ab\langle  \vb{c}_{\vb{q}_{\parallel},n}\ab(\tau)\vb{c}^{\dag}_{\vb{q}_{\parallel},n}\ab(0)\rangle,               \\
    f_{\vb{q}_{\parallel},n}\ab(\tau)
     & =-\ab\langle  \vb{c}_{\vb{q}_{\parallel},n}\ab(\tau)\,^{t}\vb{c}_{-\vb{q}_{\parallel},n}\ab(0)\rangle,               \\
    \bar{f}_{\vb{q}_{\parallel},n}\ab(\tau)
     & =-\ab\langle  \,^{t}\vb{c}^{\dag}_{-\vb{q}_{\parallel},n}\ab(\tau)\vb{c}^{\dag}_{\vb{q}_{\parallel},n}\ab(0)\rangle, \\
    \bar{g}_{\vb{q}_{\parallel},n}\ab(\tau)
     & =-\ab\langle  \,^{t}\vb{c}^{\dag}_{-\vb{q}_{\parallel},n}\ab(\tau)\,^{t}\vb{c}_{-\vb{q}_{\parallel},n}\ab(0)\rangle.
  \end{align}
  Calculating the magnon self energy, the surface dynamic spin susceptibility in bosonic Matsubara representation $\chi_{\vb{q}_{\parallel}}\ab(i\omega_{m})$ can be written as
    \begin{align}
      &\chi_{\vb{q}_{\parallel}}\ab(i\omega_{m})
      =-\frac{1}{\beta}\sum\limits_{\vb{p}_{\parallel}}\sum\limits_{i\epsilon_{m}}\bigg(\notag\\
      &\quad\ab(g_{\vb{p}_{\parallel},0}\ab(i\epsilon_{m}))_{\uparrow\uparrow}\ab(g_{\vb{p}_{\parallel}+\vb{q}_{\parallel},0}\ab(i\epsilon_{m}+i\omega_{m}))_{\downarrow\downarrow}\notag\\
      &\qquad-\ab(\bar{f}_{\vb{p}_{\parallel},0}\ab(i\epsilon_{m}))_{\downarrow\uparrow}\ab(f_{\vb{p}_{\parallel}+\vb{q}_{\parallel},0}\ab(i\epsilon_{m}+i\omega_{m}))_{\downarrow\uparrow}
      \bigg).
    \end{align}
  By analytical continuation: $i\omega_{m}\rightarrow\omega+i\delta$, we finally get the retarded surface dynamic spin susceptibility (see Appendix~\ref{deriveSS} for details). 
  In the following, we focus on the case of a rough interface and only calculate
  \begin{align}
    \delta\alpha_{\mathrm{G}}
    &=
    \frac{2S_{0}}{\hbar\omega}\frac{J^{2}_{2}}{N_{\mathrm{FI}}}\, \mathrm{Im}\, \ab(\frac{1}{N_{\parallel}}\sum_{\vb{q}_{\parallel}}\chi^{R}_{\vb{q}_{\parallel}}\ab(\omega))
    \notag
    \\
    &\equiv
    \frac{2S_{0}}{\hbar\omega}\frac{J^{2}_{2}}{N_{\mathrm{FI}}}\, \mathrm{Im}\, \chi_{\mathrm{loc}}\ab(\omega).
  \end{align}
  For the $\ab(010)$ surface, $\chi_{\mathrm{loc}}(\omega)$ can be recast as
    \begin{align}\label{chi010}
      &\mathrm{Im}\chi_{\mathrm{loc}}\ab(\omega)
        =\frac{1}{N_{\parallel}}\sum\limits_{q_{010},q^{\prime}_{010}}\int\frac{dE}{\pi}\ab(n_{\mathrm{F}}\ab(E)-n_{\mathrm{F}}\ab(E+\hbar\omega))\notag                                                                                                                                                                                          \\
       & \quad\times\bigg(\mathrm{Im}\ab(g^{R}_{q_{010},0}\ab(E))_{\uparrow\uparrow}\mathrm{Im}\ab(g^{R}_{q^{\prime}_{010},0}\ab(E+\hbar\omega))_{\downarrow\downarrow}\notag\\
       &\qquad-\mathrm{Im}\ab(f^{R}_{q_{010},0}\ab(E))_{\uparrow\downarrow}\mathrm{Im}\ab(f^{R}_{q^{\prime}_{010},0}\ab(E+\hbar\omega))_{\uparrow\downarrow}\bigg).
    \end{align}
  In contrast, for the $\ab(110)$ surface, the off-diagonal matrices of the hopping Hamiltonian in Eq.~\eqref{110t} change sign under the inversion of $q_{110}$, leading to the symmetry of the Hamiltonian $H(-q_{110}) = \tau_z H(q_{110}) \tau_z$ and of the Green's function $
 \tilde g_{-q_{110},n_{110}}(E)
 = \tau_z \tilde g_{q_{110},n_{110}}(E) \tau_z$, then we find
\begin{align}
 g_{q_{110},n_{110}} = g_{-q_{110},n_{110}},
 \
 f_{q_{110},n_{110}} = -f_{-q_{110},n_{110}}.
\end{align}
The anomalous Green's function on the (110) surface is odd with respect to $q_{110}$ and consequently the anomalous contribution to $\chi_{\mathrm{loc}}(\omega)$ vanishes.
As a result, $\chi_{\mathrm{loc}}(\omega)$ reduces to
\begin{align}\label{chi110}
    \mathrm{Im}\chi_{\mathrm{loc}}\ab(\omega)
    &=\frac{1}{N_{\parallel}}\sum\limits_{q_{110},q^{\prime}_{110}}\int\frac{dE}{\pi}\ab(n_{\mathrm{F}}\ab(E)-n_{\mathrm{F}}\ab(E+\hbar\omega))\notag\\
    &\quad\times\mathrm{Im}\ab(g^{R}_{q_{110},0}\ab(E))_{\uparrow\uparrow}\mathrm{Im}\ab(g^{R}_{q^{\prime}_{110},0}\ab(E+\hbar\omega))_{\downarrow\downarrow}.
\end{align}

\subsection{FMR Modulation in Bulk--Edge Coexistence}
We assume a weak-coupling $d$-wave superconductor and use $\Delta_{0}=2.14k_{\mathrm{B}}T_{\mathrm{c}}$ \cite{PhysRevB.76.064512} and calculate the Gilbert damping constant enhancement for the $\ab(010)/\ab(110)$ surface with the value set $a=1$, $t/\Delta_{0}=\mu/\Delta_{0}=100$, and $\delta/\Delta_{0}=0.01$.

\begin{figure}
  \centering
  \includegraphics[width=\columnwidth]{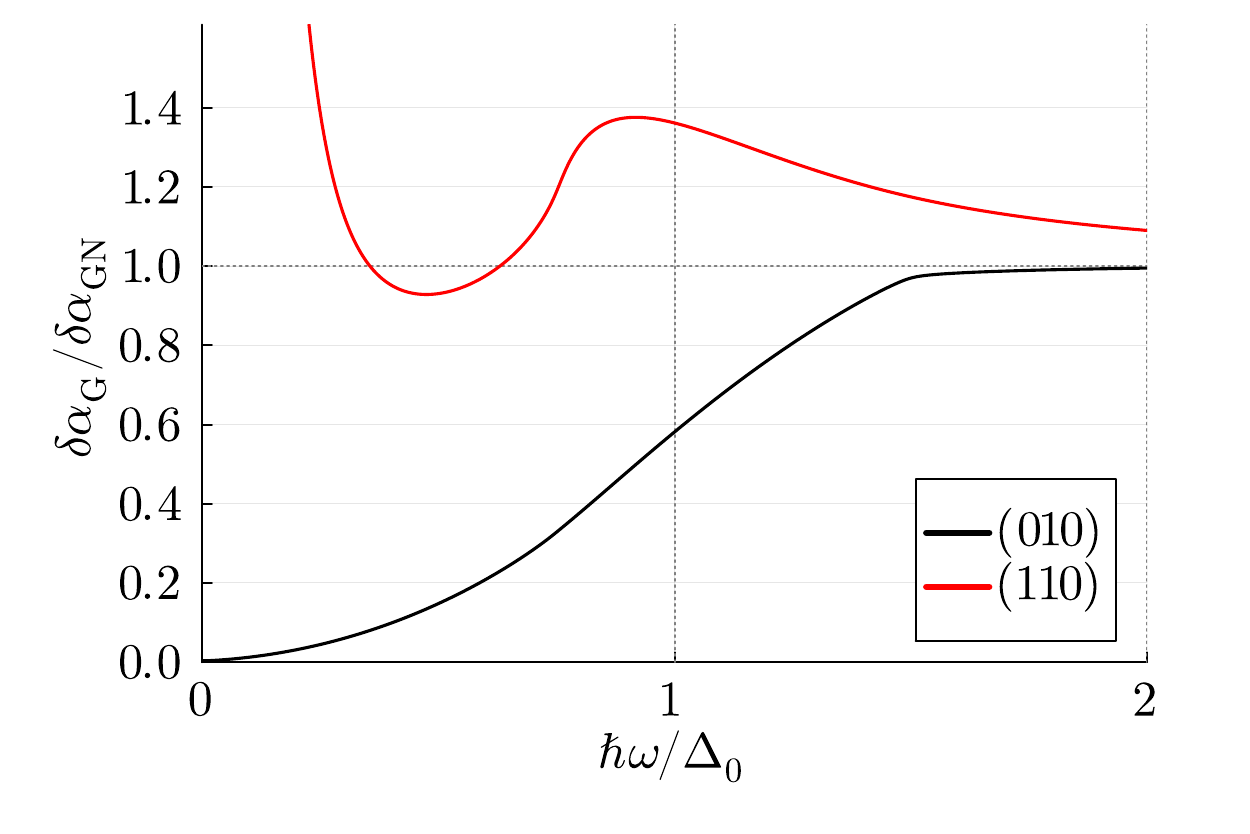}
  \caption{The frequency dependence of the enhancement of the Gilbert damping constant at $T=0$ for the $\ab(010)$ and $\ab(110)$ surface. 
  $\delta\alpha_{\mathrm{GN}}\equiv\delta\alpha_{\mathrm{G}}\ab(\omega\rightarrow0, T=T_{\mathrm{c}})$ is the enhancement of the Gilbert damping constant in the normal state at the low frequency limit.}
  \label{GDC}
\end{figure}

Fig.~\ref{GDC} shows the frequency dependence at $T=0$.
On the $\ab(010)$ surface, $\delta\alpha_{\mathrm{G}}$ is almost proportional to $\omega^{2}$ within $\hbar\omega \ll 2\Delta_{0}$, qualitatively the same as in Ref.~\cite{PhysRevB.105.205406}, although a different surface $\ab(001)$ is considered. On the other hand, there are two characteristic peaks for the $\ab(110)$ surface: a pronounced edge-to-edge excitation peak appearing in $\hbar\omega\sim0$ and an additional edge-to-bulk excitation peak emerging in $\hbar\omega\sim\Delta_{0}$.
The edge-to-edge peak is estimated $\delta\alpha_{\mathrm{G}}\sim\delta\alpha_{\mathrm{GN}}\times10^{3}$ and linewidth corresponds to phenomenological broadening $\delta$, which encodes the finite quasiparticle lifetime $\tau$.
This excitation originates from ABS at zero energy and is followed by a decay of the power law of $\omega^{-3}$.

\begin{figure}
  \centering

  \begin{minipage}{0.48\linewidth}
    \subcaption{}
    \includegraphics[width=\columnwidth]{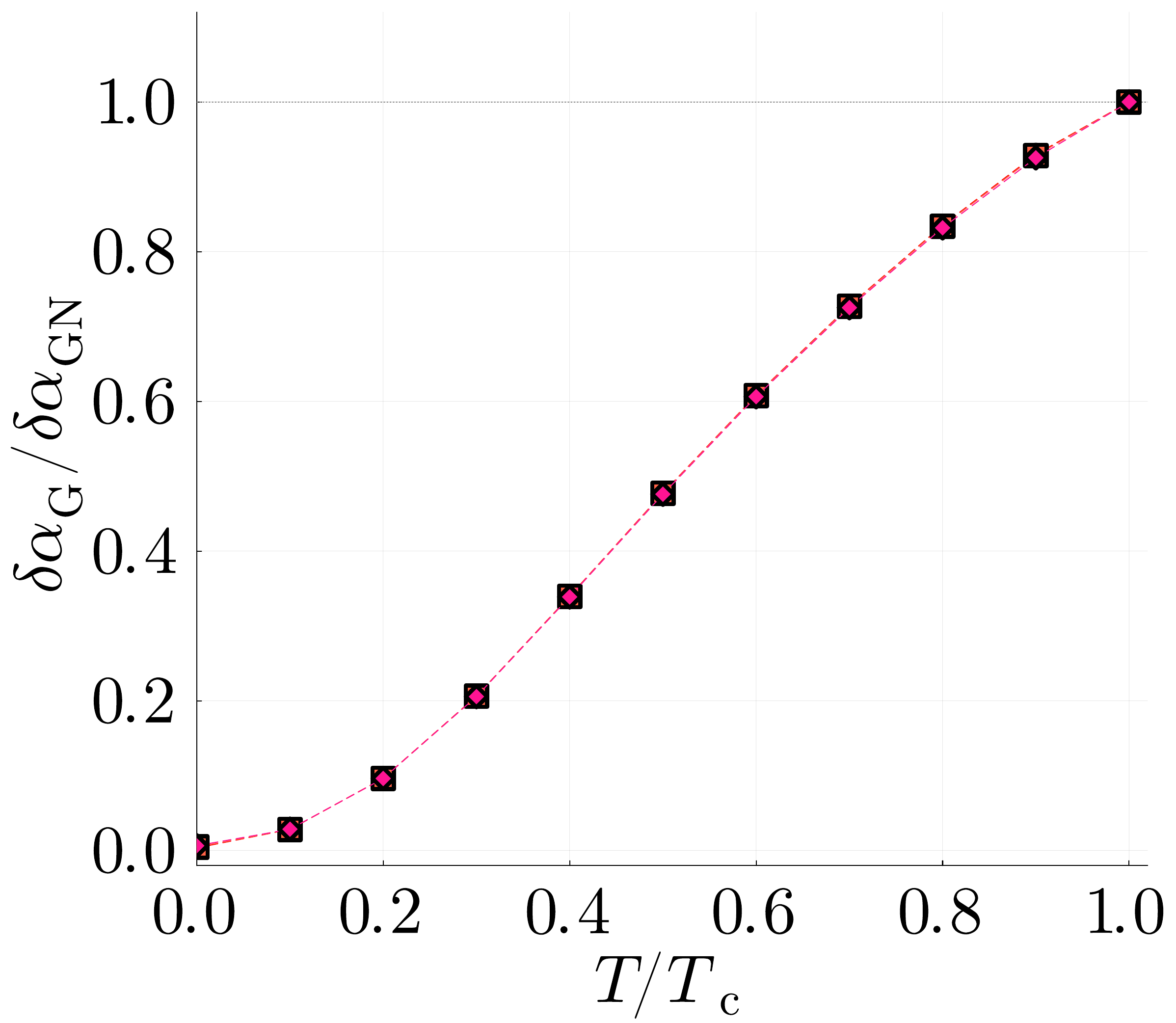}
    \label{010temp_low}
  \end{minipage}
  \begin{minipage}{0.48\linewidth}
    \subcaption{}
    \includegraphics[width=\columnwidth]{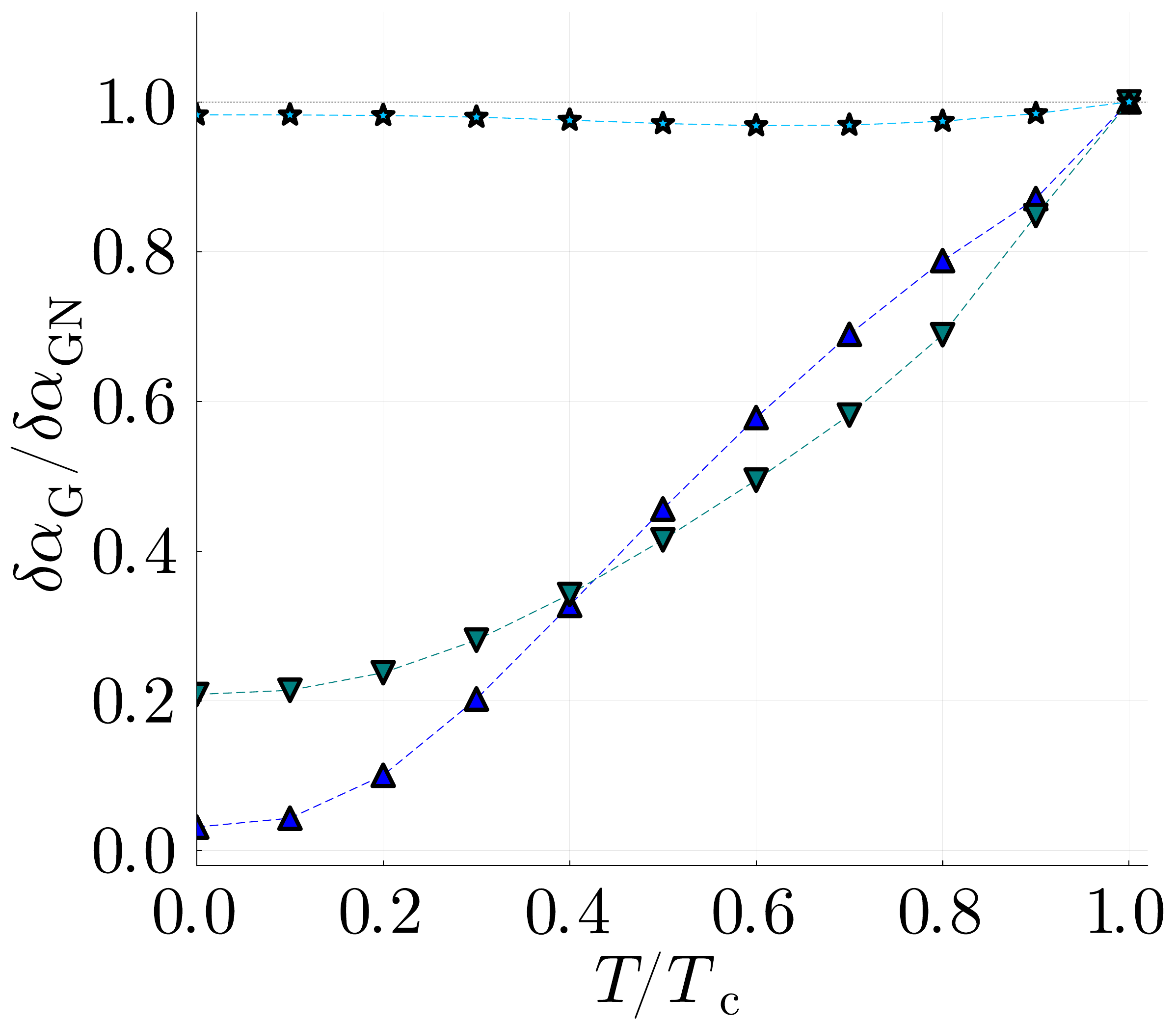}
    \label{010temp_high}
  \end{minipage}
  \begin{minipage}{0.48\linewidth}
    \subcaption{}
    \includegraphics[width=\linewidth]{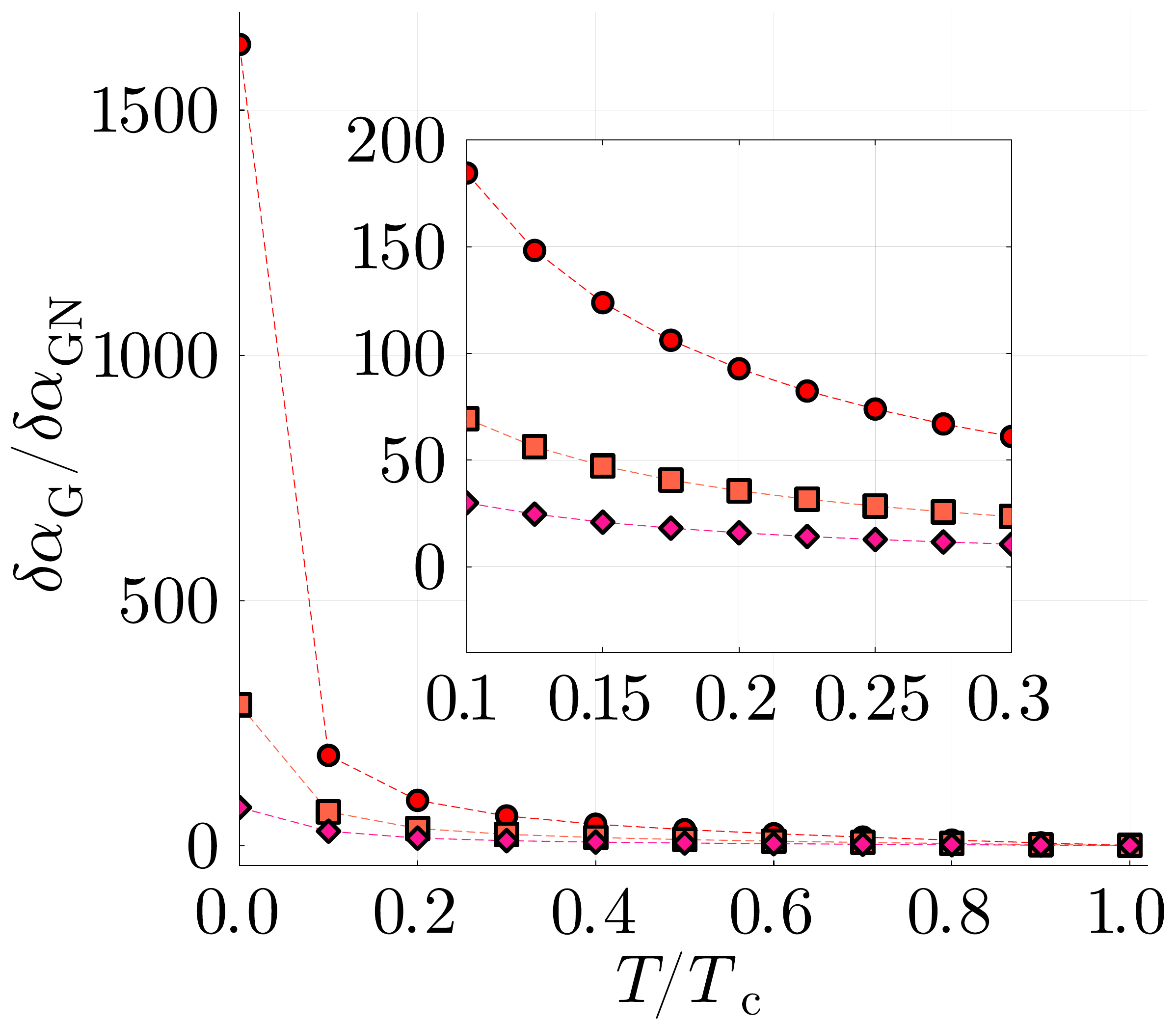}
    \label{110temp_low}
  \end{minipage}
  \begin{minipage}{0.48\linewidth}
    \subcaption{}
    \includegraphics[width=\linewidth]{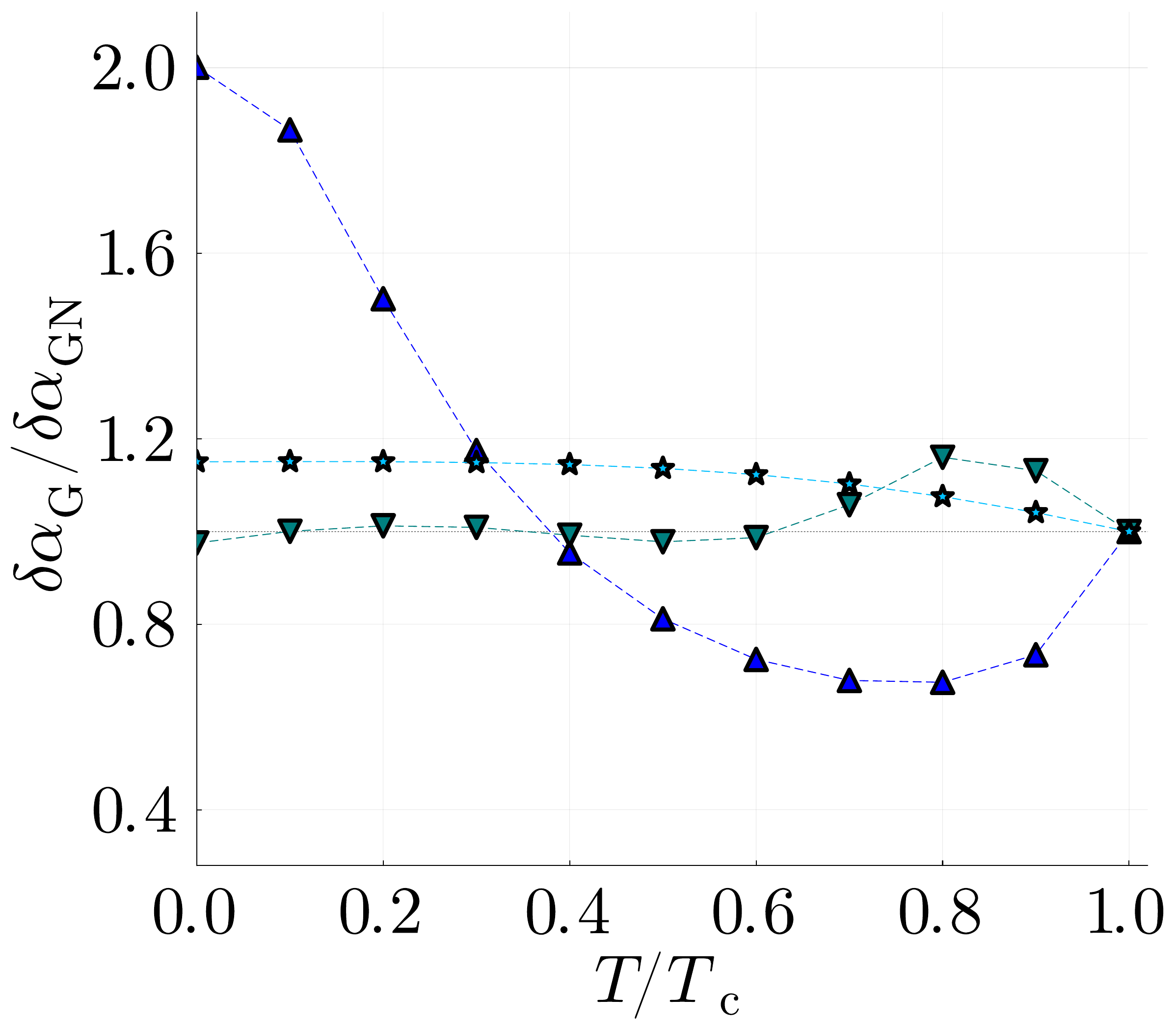}
    \label{110temp_high}
  \end{minipage}
  \begin{minipage}{0.48\linewidth}
    \includegraphics[width=\linewidth]{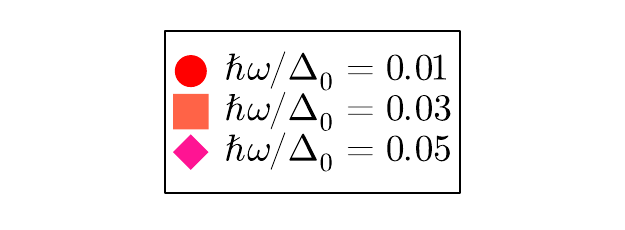}
  \end{minipage}
  \begin{minipage}{0.48\linewidth}
    \includegraphics[width=\linewidth]{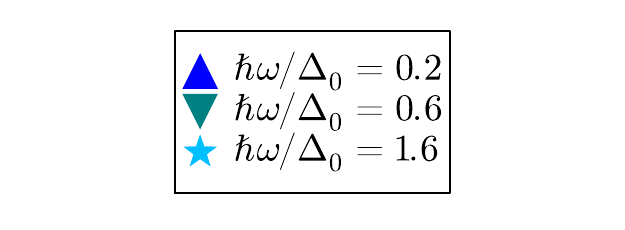}
  \end{minipage}
  \caption{Temperature dependence of the enhancement of the Gilbert damping constant for the $\ab(010)$ [(a), (b)] and $\ab(110)$ [(c), (d)] surfaces. Panels (a) and (c) show the low-energy regime, while (b) and (d) correspond to the intermediate- and high-energy regimes.}
  \label{tempdep}
\end{figure}

Fig.~\ref{tempdep} provides the temperature dependence in low-, intermediate-, and high-energy regimes.
In Fig.~\ref{010temp_low}, the three curves for $\delta\alpha_{\mathrm{G}}$ corresponding to the low-energy regime nearly overlap ($\hbar\omega/\Delta_0=0.01, 0.03, \mathrm{and}\, 0.05$). They also approximately scale as $T^{2}$ on the $\ab(010)$ surface, consistent with Ref.~\cite{PhysRevB.105.205406}, despite the different surface $\ab(001)$ is considered there. Moreover, for the $\ab(110)$ surface, $\delta\alpha_{\mathrm{G}}$ decreases as $T^{-1}$ at low temperatures and decreases exponentially at intermediate temperatures in the low frequency limit, originating from the contribution of the edge states (see Fig.~\ref{110temp_low}).
In Fig.~\ref{110temp_high}, the green line, which corresponds to $\hbar\omega/\Delta_{0}=0.6$, shows a local peak at high temperatures $T \sim 0.8 T_{\rm c}$, where the gap amplitude is $E_{\rm g} \sim 0.6 \Delta_0$. 
This is the result of the capture of an edge-to-bulk excitation peak at the gap frequency $\hbar\omega \sim E_{\rm g}$.

We also note that these scalings of edge-to-edge excitation can be understood from the density of states of the ABS, which is given by the Lorentzian:
\begin{equation}
    -\frac{1}{N_\parallel}
    \sum_{q_{110}} 
    \mathrm{Im}\ab(g^{R}_{q_{110},0}\ab(E))_{\uparrow\uparrow/\downarrow\downarrow}
    \sim
    \frac{\delta}{E^{2}+\delta^{2}}.
\end{equation}
For the frequency scaling at $T=0$, the enhancement in the limit $\delta\ll\hbar\omega$ is evaluated by replacing one of the two Lorentzians with a delta function, leading to
\begin{align}
    \delta\alpha_{\mathrm{G}} 
    &\propto\frac{\mathrm{Im}\chi_{\rm loc}(\omega)}
    {\omega}
    \notag
    \\
    &\propto 
    \frac{1}{\omega}
    \int_{-\hbar\omega}^{0}dE\delta(E)\frac{\delta}{(E+\hbar\omega)^{2}+\delta^{2}}
    \notag\\&
    \propto\frac{\delta}{\omega^3}.
\end{align}
In the opposite limit $\delta\gg\hbar\omega$, the Lorentzians are replaced by $1/\delta$, so that the divergence is cutoff by $\delta$.
For the temperature scaling in the low frequency regime, the power-law decay $T^{-1}$ can be explained by the difference of the Fermi distribution functions, which yields a factor proportional to $\omega/T$.
These scalings are valid under the assumption that $\delta$ is a constant broadening , where its frequency and temperature dependence can be neglected.
In contrast, from theoretical perspectives, when impurity scattering strongly influences $d$-wave superconductors, the quasiparticle lifetime $\tau$ acquires temperature dependence through the self energy $\Sigma_{\mathrm{SC}}(T)$, both in the bulk~\cite{PhysRevB.72.214512, PhysRevB.77.184504} and for surface ABS~\cite{PhysRevB.78.104524}.
In addition, it has also been experimentally reported that the quasiparticle lifetime exhibits temperature dependence~\cite{FASANO2018798}.
Based on these previous results, $\delta$ can show the temperature dependence $\delta=\delta\ab(T)$ due to the impurity scattering and modify the damping scalings.

\subsection{Discussion and Experimental Implications}
In the low-frequency limit near the transition temperature, a coherence peak is not observed in the $d$-wave case, as shown in Fig.~\ref{tempdep}, whereas a coherence peak appears in the $s$-wave case \cite{
PhysRevB.99.144411, PhysRevB.105.205406}.
This absence is due to the quasiparticle bulk node, which suppresses the divergence of the density of state at the gap edge, as discussed in Ref.~\cite{PhysRevB.105.205406}.

The low-frequency edge-to-edge excitation provides a direct signature of ABS. 
In particular, the enhancement of $\delta\alpha_{\mathrm{G}}$ in the limit $\hbar\omega \to 0$ indicates that ABS make a positive contribution to the Gilbert-damping correction. 
This interpretation is consistent with previous FMR experiments on Py/YBCO junctions \cite{PhysRevB.104.144428}, as well as with theoretical studies of proximity-induced ABS in FI/N/superconductor heterostructures \cite{PhysRevB.103.L100406, Silaev2020-bv}.

It is also important to assess additional interfacial couplings beyond the spin-transfer term in Eq.~\eqref{Hint}. 
In particular, the exchange-bias contribution \cite{PhysRevB.105.205406},
\begin{equation}
  H_{\mathrm{ex}}=\sum_{\vb{k},\vb{q}_{\parallel}}J_{\vb{k},\vb{q}_{\parallel}}S^{z}_{\vb{k}}s^{z}_{\vb{q}_{\parallel}},
\end{equation}
can modify the ABS spectrum. 
This term can be decomposed into a uniform Zeeman contribution $H_{\mathrm{Z}}$ and a disorder contribution $H_{\mathrm{dis}}$ localized in the outermost surface layer. 
The Zeeman term can gap the edge states and thereby suppress the edge-to-edge peak in $\delta\alpha_{\mathrm{G}}$ near $\omega \sim 0$ \cite{PhysRevB.107.144504}. 
The magnitude of this effect is controlled by the ratio of the interfacial exchange scale to the superconducting gap. 
Using the typical estimate $J_{1}/\sqrt{N_{\mathrm{FI}}}\sim 0.1\,\mathrm{meV}$ \cite{nogues1999exchange, PhysRevB.105.205406}, we find that $H_{\mathrm{Z}}/\Delta_{0}\lesssim 10^{-2}$ for cuprate superconductors with relatively high $T_{\mathrm{c}}$ \cite{RevModPhys.60.585}, so that the Zeeman-induced gap should be negligible. 
For heavy-fermion superconductors with $T_{\mathrm{c}}\sim 1\,\mathrm{K}$ \cite{petrovic2001heavy, Zhou_2013}, by contrast, $H_{\mathrm{Z}}/\Delta_{0}$ can reach $O(10^{-1})$, although it still remains below unity for the parameters considered here. 
In that case, the ABS contribution is expected to be reduced, but not completely suppressed, unlike the parameter regime discussed in Ref.~\cite{PhysRevB.107.144504}.

The disorder term $H_{\mathrm{dis}}$ can also affect ABS, as discussed in Refs.~\cite{PhysRevB.65.064522, PhysRevB.70.184505, PhysRevB.98.134516}. 
While its first-order contribution vanishes, second- and higher-order processes renormalize the surface Green's function and may shift or broaden the ABS-related features.

Finally, our results suggest an experimentally relevant route for detecting ABS. 
In both cuprate and heavy-fermion superconductors, an edge-to-edge excitation should be observable as a low-frequency enhancement of $\delta\alpha_{\mathrm{G}}$, indicating that FMR can serve as a spin-sensitive probe of ABS. 
This approach is complementary to tunneling spectroscopy, where ABS is typically identified through a zero-bias conductance peak \cite{PhysRevLett.74.3451, kashiwaya2000tunnelling}. 
Moreover, for superconductors with $T_{\mathrm{c}}\sim 1\,\mathrm{K}$, the edge-to-bulk excitation peak may also fall within an experimentally accessible frequency window. This is because FMR modulation can probe the GHz range, which corresponds roughly to the 1K energy scale $\ab(30\,\mathrm{GHz}\approx 1\,\mathrm{K})$, providing information that is less directly accessible in conventional conductance measurements.
\section{Conclusion}\label{conclusion}
In this paper, we have extended the theory of FMR modulation that can treat bulk--boundary coexistence in topological materials. We consider a system consisting of an FI and an arbitrary semi-infinite FS and show that the magnon self energy is given by the surface dynamical spin susceptibility of the FS. 

We have also calculated the frequency and temperature dependences of the enhanced Gilbert damping constant $\delta\alpha_{\mathrm{G}}$ for the $\ab(110)$ surface of a $d$-wave superconductor. We reveal that $\delta\alpha_{\mathrm{G}}$ has a pronounced peak that appears in $\hbar\omega\sim0$, followed by a decay of the power law $\omega^{-3}$ and an additional peak that emerges in $\hbar\omega\sim\Delta_{0}$. The first peak represents an edge-to-edge excitation, arising from quasiparticle excitation within the edge states, and the second peak originates from an edge-to-bulk excitation, associated with quasiparticle excitation from edge to bulk states. 
Furthermore, $\delta\alpha_{\mathrm{G}}$ decreases as $T^{-1}$ at low temperatures and decreases exponentially at intermediate temperatures in the low frequency limit.

Our results demonstrate that the present formalism provides a unified framework for analyzing FMR modulation in systems with bulk–boundary coexistence. Because the surface dynamical spin susceptibility is computed from the surface Green’s function of an arbitrary semi-infinite fermionic system, the formulation naturally incorporates both bulk and boundary states on the same footing. 
It is broadly applicable to multiband tight-binding models beyond the $d$-wave superconductor studied here. It should therefore provide an efficient route to quantitative calculations of FMR modulation in a wide class of topological materials, such as nodal topological superconductors and topological semimetals, and more generally in systems where low-energy bulk and boundary states coexist.
\appendix

\section{Recursive Green's function method}\label{RGFcal}
The surface Green's function is the solution to the following equation:
\begin{align}
  g^{R}_{\vb{q}_{\parallel},0}\ab(E)
   & =\ab(E+i\delta-\vb{u}-\vb{t}^{\dag}g^{R}_{\vb{q}_{\parallel},0}\ab(E)\vb{t})^{-1}
  \notag
  \\&
  \equiv X_\bullet g^{R}_{\vb{q}_{\parallel},0}\ab(E),
\end{align}
with the $2N \times 2N$ matrix
\begin{align}
  X = \begin{pmatrix}
        0            & \vb{t}^{-1}                   \\
        -\vb{t}^\dag & (E+i\delta-\vb{u})\vb{t}^{-1}
      \end{pmatrix},
\end{align}
where we define the M\"obius transformation as
\begin{align}
  \begin{pmatrix}
    a & b
    \\
    c & d
  \end{pmatrix}_\bullet
  Y
   & =
  (aY+b)(cY+d)^{-1},
\end{align}
which is associative, $X_\bullet (Y_\bullet Z) = (XY)_\bullet Z$.
The solution is given in closed form
\begin{align}
  g^{R}_{\vb{q}_{\parallel},0}\ab(E) = {X^\infty}_\bullet 0.
\end{align}
Note that the solution is independent of the initial value; therefore, we set $0$ in the equation above.
We also define the matrix $\vb{O}$:
\begin{equation}
  \vb{O}=
  \begin{pmatrix}
    \vb{O}_{11} & \vb{O}_{12} \\\vb{O}_{21}&\vb{O}_{22}
  \end{pmatrix},
\end{equation}
which diagonalizes $X$ as $\vb{O}^{-1}X\vb{O}=\mathrm{diag}\ab(\lambda_{1},\lambda_{2},\cdots,\lambda_{2N})$ and $|\lambda_{1}|<|\lambda_{2}|<\cdots<|\lambda_{2N}|$. Thus, the surface Green's function is written as
\begin{equation}
  g^{R}_{\vb{q}_{\parallel},0}(E)=\vb{O}_{12}\vb{O}^{-1}_{22}.
\end{equation}
The surface Green's function in semi-infinite space is obtained by diagonalizing the $2N\times2N$ matrix $X$, and the computational cost is comparable to that of evaluating the bulk Green's function, when we only consider the nearest-neighbor hopping.

\section{Effect the surface roughness on magnon self energy}
\label{conf.ave.}
We consider the effect of surface roughness as configuration average introduced in Ref.~\cite{PhysRevB.105.205406, PhysRevB.106.L161406, Ominato_2025}.
By defining $J_{1}$ and $J^{2}_{2}$ as the mean and variance of the interface coupling strength, respectively, the in-plane coupling constant satisfies the following equations:
\begin{align}
  \frac{1}{2}\overline{\sum\limits_{i}J_{i,j_{\parallel}}}
   & =J_{1},\quad        \\
  \frac{1}{4}\overline{\sum\limits_{i,i^{\prime}}J_{i,j_{\parallel}}J_{i^{\prime},j^{\prime}_{\parallel}}}
   & =J^{2}_{1}+J^{2}_{2}\delta_{j_{\parallel},j^{\prime}_{\parallel}}.
\end{align}
The averaged magnon self energy is expressed as
\begin{align}
  \overline{\Sigma^{R}_{\vb{k}=0}\ab(\omega)}
  &=-\sum_{\vb{q}_{\parallel}}\overline{\frac{\ab|J_{\vb{k}=0,\vb{q}_{\parallel}}|^{2}}{4}}\chi^{R}_{\vb{q}_{\parallel}}\ab(\omega)
  \notag
  \\
  &=-\frac{J^{2}_{1}}{N_{\mathrm{FI}}}\chi^{R}_{\vb{q}_{\parallel}=0}\ab(\omega)-\frac{J^{2}_{2}}{N_{\mathrm{FI}}}\frac{1}{N_{\mathrm{\parallel}}}\sum\limits_{\vb{q}_{\parallel}}\chi^{R}_{\vb{q}_{\parallel}}\ab(\omega).
\end{align}
The first and second terms corresponds to flat and rough limit interface, respectively.

\section{Derivation of Spin Susceptibility}
\label{deriveSS}

We derive the retarded spin susceptibility $\chi_q^{\mathrm R}$, where $(q)=(\boldsymbol q_\parallel, \omega)$, in a general case.
Reducing Eq.~\eqref{eq:chi_q}, we obtain
\begin{align}
    \chi_{q_m}
     & = -\frac{1}{\beta} \sum_{p_n}
    \, \mathrm{tr} \, \biggl(g_{p_n} \frac{\sigma^+}{2} g_{p_n+q_m} \frac{\sigma^-}{2}
    \notag                                 \\ &\hspace{8em}
    -\bar{f}_{p_n} \frac{\sigma^+}{2} f_{p_n+q_m} \frac{{}^t\sigma^-}{2}
    \biggr),
\end{align}
where we use the notation $(p_n)=(\boldsymbol p_\parallel, i\epsilon_n)$ and $(q_m)=(\boldsymbol q_\parallel, i\omega_m)$.
The Matsubara sum is calculated as the following integral, together with the analytic continuation $i\omega_m \to \omega + i0$
\begin{align}
     &
    \chi_{q}^{\mathrm R}
    = \sum_{\boldsymbol p_\parallel} \int d\epsilon \,
    \biggl[
    \notag \\&
        n(\epsilon) \,
        \mathrm{tr} \,
        \biggl(-\mathcal A_p \frac{\sigma^+}{2} g_{p+q}^{\mathrm R} \frac{\sigma^-}{2}
        +
        \bar{\mathcal B}_p \frac{\sigma^+}{2} f_{p+q}^{\mathrm R} \frac{{}^t\sigma^-}{2}
        \biggr)
    \notag \\&+
        n(\epsilon + \omega) \,
        \mathrm{tr} \,
        \biggl(-g_{p}^{\mathrm A} \frac{\sigma^+}{2} \mathcal A_{p+q} \frac{\sigma^-}{2}
        +
        \bar{f}_{p}^{\mathrm A} \frac{\sigma^+}{2} \mathcal B_{p+q} \frac{{}^t\sigma^-}{2}
        \biggr)
        \biggr],
\end{align}
Here, we define the spectral matrices
\begin{align}
    \mathcal A_p = -\frac{g_p^{\mathrm R} - g_p^{\mathrm A}}{2\pi i},
    \
    \mathcal B_p = -\frac{f_p^{\mathrm R} - f_p^{\mathrm A}}{2\pi i},
    \
    \bar{\mathcal B}_p = -\frac{\bar{f}_p^{\mathrm R} - \bar{f}_p^{\mathrm A}}{2\pi i},
\end{align}
where $\mathrm{tr} \mathcal A_p$ coincides with the spectral function.
Note that $\mathcal A_p = \mathcal A_p^\dag$ and $\bar{\mathcal B}_p = \mathcal B_p^\dag$.

The imaginary part of the spin susceptibility is given by
\begin{align}
    \mathrm{Im} \chi_q^{\mathrm R}
     & = \sum_{\boldsymbol p_\parallel} \int d\epsilon \, \ab[n(\epsilon)-n(\epsilon+\omega)]
    \pi \,
    \mathrm{tr} \, \ab(\mathcal A_p \frac{\sigma^+}{2} \mathcal A_{p+q} \frac{\sigma^-}{2})
    \notag                                             \\&
    \quad +
    \sum_{\boldsymbol p_\parallel} \int d\epsilon \,
    \mathrm{Im} \,
    \mathrm{tr} \, \biggl[
        n(\epsilon)
        \ab(\bar{\mathcal B}_p \frac{\sigma^+}{2} f_{p+q}^{\mathrm R} \frac{{}^t\sigma^-}{2})
    \notag                                             \\&\hspace{4em}
        +
        n(\epsilon+\omega)
        \ab(\bar f_p^{\mathrm A} \frac{\sigma^+}{2} \mathcal B_{p+q} \frac{{}^t\sigma^-}{2})
        \biggr].
\end{align}
For the spin--orbit-free single-band system with spin-singlet pairing $\propto i \sigma_y \tau_y$, the Green's functions satisfy
\begin{align}
 g_p^{\mathrm R} = g_{p\uparrow\uparrow}^{\mathrm R} \sigma_0,
 \
    f_{p}^{\mathrm R/\mathrm A} = f_{p \uparrow\downarrow}^{\mathrm R/\mathrm A} i\sigma_y,
    \
    \bar f_p^{\mathrm R/\mathrm A} = - f_p^{\mathrm R/\mathrm A},
\end{align}
and the spectral matrices are given by
\begin{align}
    \mathcal A_{p} = -\frac{1}{\pi}\, \mathrm{Im} \, g_{p}^{\mathrm R},
    \
    \mathcal B_p = -\frac{1}{\pi}\, \mathrm{Im}\, f_{p}^{\mathrm R},
    \
    \bar{\mathcal B}_p = -\mathcal B_p,
\end{align}
which are real valued. 
The imaginary part of the susceptibility reduces to 
\begin{align}
    \mathrm{Im} \, \chi_q^{\mathrm R}
    &
    = \sum_{\boldsymbol p_\parallel} \int d\epsilon \,
    \ab[n(\epsilon)-n(\epsilon+\omega)]
    \notag\\&\quad\times
    \pi \, \mathrm{tr}\, \ab(\mathcal A_p \frac{\sigma^+}{2} \mathcal A_{p+q} \frac{\sigma^-}{2}
    - 
    {\mathcal B}_p \frac{\sigma^+}{2} \mathcal B_{p+q} \frac{{}^t\sigma^-}{2}).
\end{align}

\acknowledgments
This work was supported by JSPS KAKENHI for Grants (Grants Nos.~JP23H01839, JP24H00322, JP24K06951, JP24H00853, and JP25K07224), Waseda University Grant for Special Research Projects (Grants No.~2025C-651 and No.~2025R-061), and by the National Natural Science Foundation of China (NSFC) under Grant No.~12374126.

\bibliography{ref.bib}

\end{document}